\documentclass[final,authoryear,5p,times,twocolumn]{elsarticle}

 \usepackage{graphicx}

\usepackage{amssymb}

\journal{Nuclear Physics B}

\begin{document}

\begin{frontmatter}



\title{The line parameters and ratios as the physical probe
of the line emitting regions in AGN}


\author[doa]{D. Ili\'c}
\ead{dilic@matf.bg.ac.rs}
\author[aob]{J. Kova\v cevi\'c}
\author[aob]{L. \v C. Popovi\'c}
\address[doa]{Department of Astronomy, Faculty of Mathematics,
University of Belgrade, Studentski trg 16, 11000 Belgrade, Serbia}
\address[aob]{Astronomical Observatory, Volgina 7, 11060
Belgrade, Serbia}

\begin{abstract}
Here we discuss the physical conditions in the emission line regions
(ELR) of active galactic nuclei (AGN), with the special emphasize on
the unresolved problems, e.g. the stratification of the Broad Line
Region (BLR) or the failure of the photoionization to explain the
strong observed optical Fe II emission. We use here different line
fluxes in order to probe the properties of the ELR, such as the
hydrogen Balmer lines (H$\alpha$ to H$\varepsilon$), the helium
lines from two subsequent ionization levels (He II $\lambda$4686 and
He I $\lambda$5876) and the strongest Fe II lines in the wavelength
interval $\lambda\lambda 4400-5400 \, \AA$. We found that the
hydrogen Balmer and helium lines can be used for the estimates of
the physical parameters of the BLR, and we show that the Fe II
emission is mostly emitted from an intermediate line region (ILR),
that is located further away from the central continuum source than
the BLR.

\end{abstract}


\end{frontmatter}


\section{Introduction}

Active galactic nuclei (AGN) are a ubiquitous phenomena, in a sense
that most galaxies experience some sort of activity in their nucleus
during their evolution. The most accepted scenario of the AGN
structure is that they are powered by the accretion of matter onto a
supermassive black hole (SMBH). One of the ways to study the inner
emitting regions of an AGN, is by analyzing its emission lines, i.e.
the broad (BELs) and narrow emission lines (NELs). So far many
papers and textbooks are devoted to the physical properties of the
emission line regions (ELR) \citep[see e.g.][and references
therein]{BG92,Su00,OF06}, but, there are still many open issues.

Spectroscopy, in general, offers different methods for diagnostics
of the emitting plasma \citep[see e.g.][]{Gr97,OF06}, but these
methods could not be properly used to probe the physical conditions
in some ELR of AGN, as e.g. in the Broad Line Region (BLR), since
the forbidden lines, that are usually employed in plasma diagnostics
of the Narrow Line Region (NLR) or HII regions, are not present in
the BELs spectrum. Particularly, it is difficult to find a direct
method which would only use the observed BELs to determine the
temperature and density in the BLR. On the other hand, the optical
Fe II ($\lambda\lambda$ 4400-5400 \AA) lines are one of the most
interesting features in the AGN spectrum. The origin of the optical
Fe II extreme emission and the geometrical place of the Fe II
emission region in AGN, are still open questions. Also, there are
many correlations of the Fe II lines and other AGN spectral
properties which need physical explanation such as: EW Fe II vs.
$\frac{\mathrm{EW [O III]}}{\mathrm{EW H\beta}}$, EW Fe II vs. peak
[O III], EW Fe II and FWHM of H$\beta$, etc. \citep{BG92}.

In order to estimate the physical conditions (such as the
temperature and hydrogen density) of the BLR we use the Balmer and
helium line ratios obtained in two ways: (i) using the
photoionization code CLOUDY, a spectral synthesis code designed to
simulate conditions within a plasma and model the resulting
spectrum, and (ii) extracting a sample of AGN from the Sloan Digital
Sky Survey (SDSS) database. We investigate these line ratios in
order to find conditions in the BLR where so-called Boltzmann-plot
(BP) method is applicable \citep{Gr97,Po03,Po06a}. For these special
cases, we study the correlations between the average temperature,
hydrogen density and He II/He I line ratio. Moreover, we present an
investigation of the optical Fe II emission in AGN, for which we
have used an additional sample of 111 AGN from the SDSS database.
The strongest Fe II lines are identified and classified into four
groups according to the lower level of the transition: $^4$F, $^6$S,
$^4$G and $^2$D1. In this progress report, we report our recent
investigations of the physical and kinematical properties of the BLR
and Fe II emitting region. This report is organized as follows: in
\S 2 we describe the numerical simulations of the BLR and briefly
introduce the BP method, and give the analysis of the simulated
BELs; in \S 3 we study the SDSS sample of hydrogen Balmer and helium
lines, while in \S 4 the selection and analysis of the SDSS sample
of Fe II lines are given; in \S 5 we discuss some results and
finally, in \S 6 our conclusions are given.


\section{The BEL simulations}

In order to study the BELs, we simulated the BLR emission line
spectrum from different grids of the BLR photoionization models
using the CLOUDY code \citep[version C07.02.01:][]{Fe98, Fe06}.
Input parameters for the simulations are chosen to match the
standard conditions in the BLR \citep{Fe06, KG00, KG04}, i.e. the
solar chemical abundances, the constant hydrogen density, the code's
AGN template for the incident continuum shape. We compute an
emission-line spectrum for the coordinate pair of hydrogen gas
density $n_{\rm H} [{\rm cm^{-3}}]$ and hydrogen-ionizing photon
flux $\Phi_{\rm H} [{\rm cm^{-2}s^{-1}}]$. The grid dimensions
spanned 4 orders of magnitude in each direction, and with an origin
of log $n_{\rm H} = 8$, log $ \Phi_{\rm H} = 17$ was stepped in 0.2
dex increments. A column density $N_{\rm H} [{\rm cm}^{-2}]$ was
kept constant in producing one grid of simulations. Even though many
authors claim that the most probable value of the BLR column density
is $N_{\rm H} = 10^{23} {\rm cm}^{-2}$ \citep{Du98, KG00, KG04}, we
produce here 5 different grids of models changing the column density
in the range $N_{\rm H} = 10^{21} - 10^{25}{\rm cm}^{-2}$.

We further analyze the BEL fluxes\footnote{The CLOUDY code gives all
line fluxes normalized to the H$\beta$ flux. Since it has no
influence in our analysis, we have used these values.} from the
CLOUDY grids of models. We consider in our analysis the hydrogen
Balmer lines (H$\alpha$ to H$\varepsilon$) and the flux ratio $R$ of
the helium lines He II $\lambda$4686 and He I $\lambda$5876, defined
as $R=F({\rm He II} \lambda 4686)/F({\rm He I} \lambda 5876)$, where
$F$ is the line flux. We particularly consider these helium lines,
since they are the lines of the same element but in two different
ionization states, thus their ratio He II $\lambda$4686/He I
$\lambda$5876 is sensitive to the change in the temperature and
density \citep[see e.g.][]{Gr97}. Besides, these lines are in the
same spectral range as Balmer lines. Additionally, from the grids of
models we consider in our analysis an averaged temperature, which is
the electron temperature averaged over the BLR radius ($T_{\rm av}$
further in the text).

\begin{figure}
\centering
\includegraphics[width=0.22\textwidth,angle=-90]{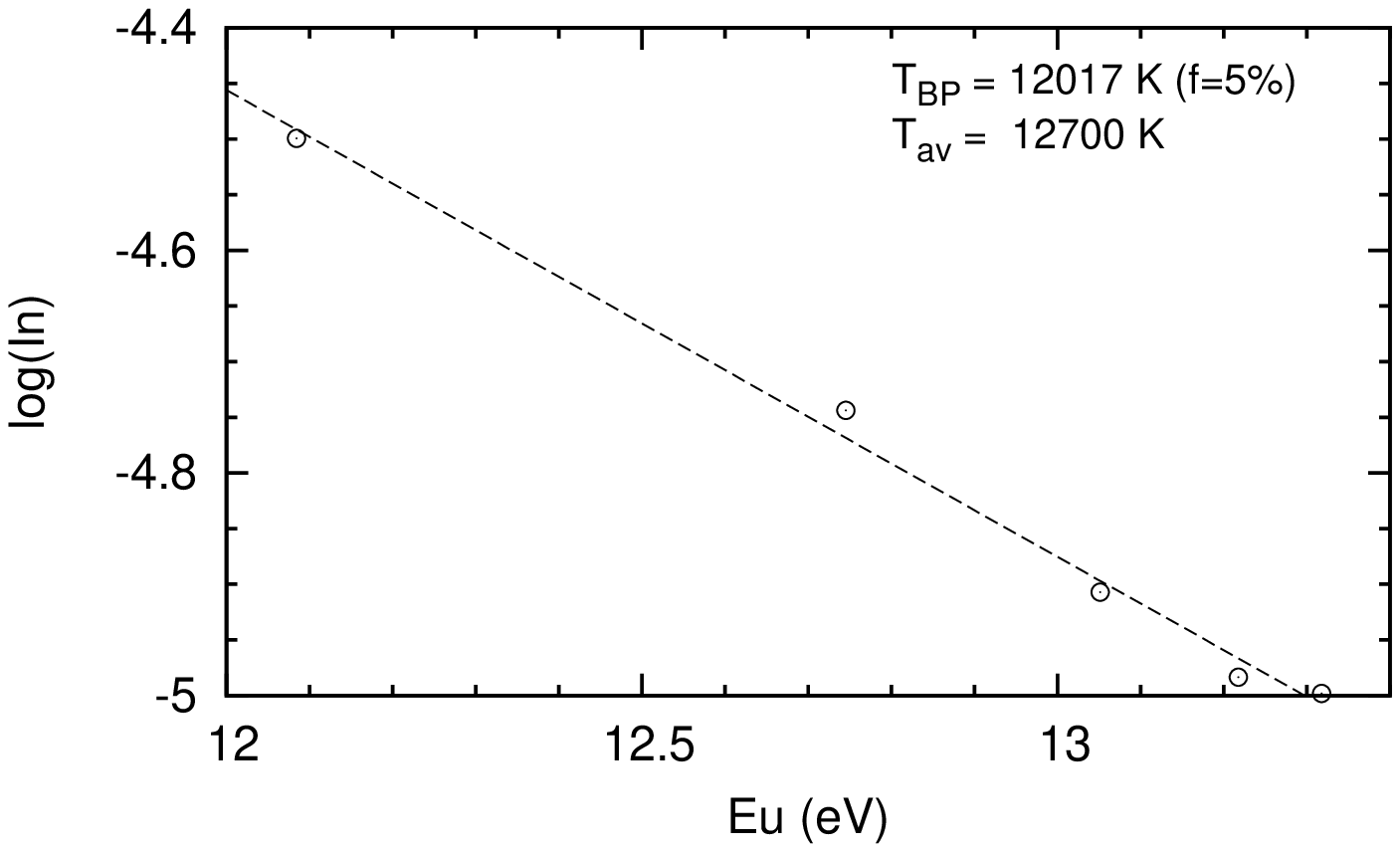}
\includegraphics[width=0.22\textwidth,angle=-90]{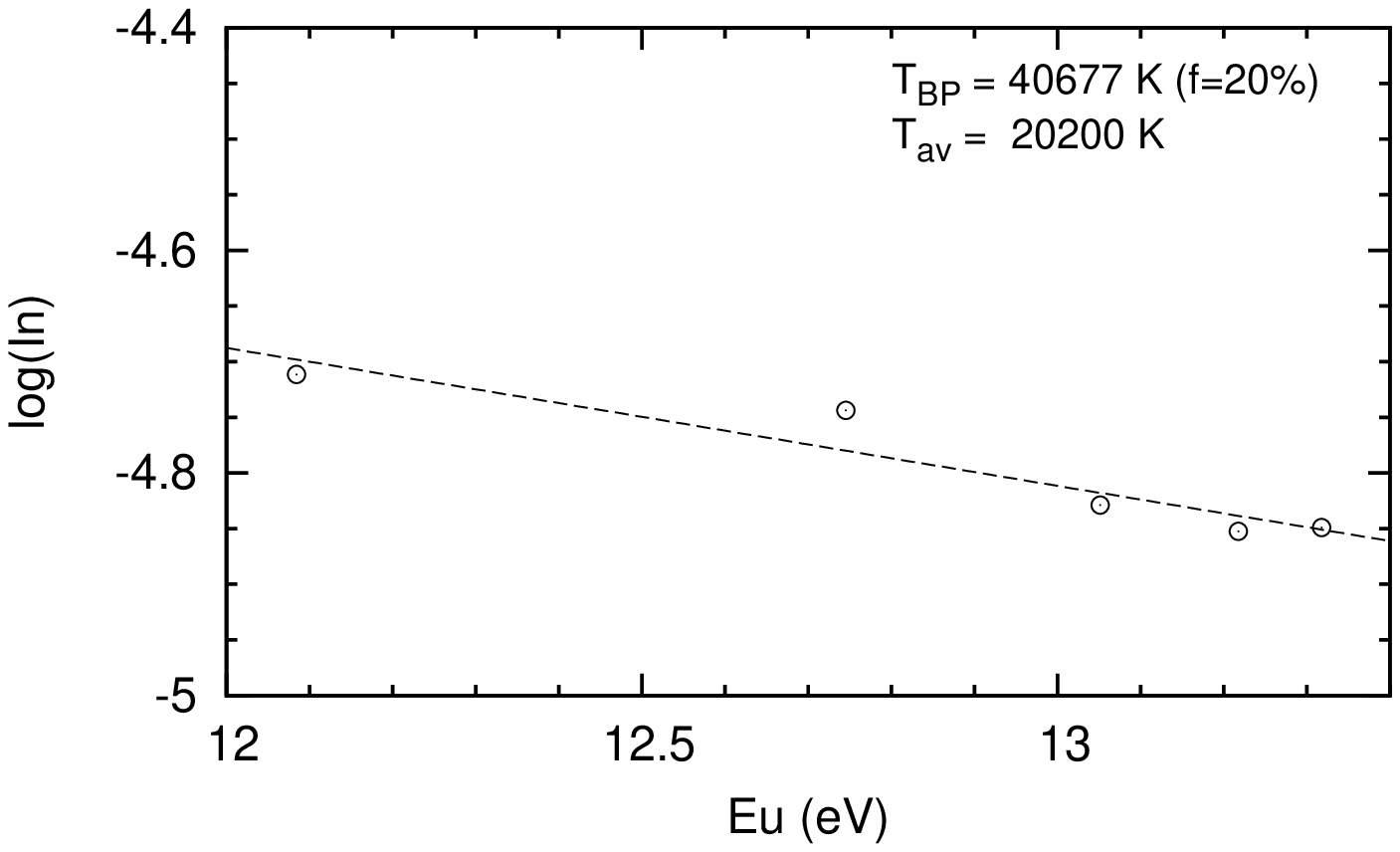}
\includegraphics[width=0.22\textwidth,angle=-90]{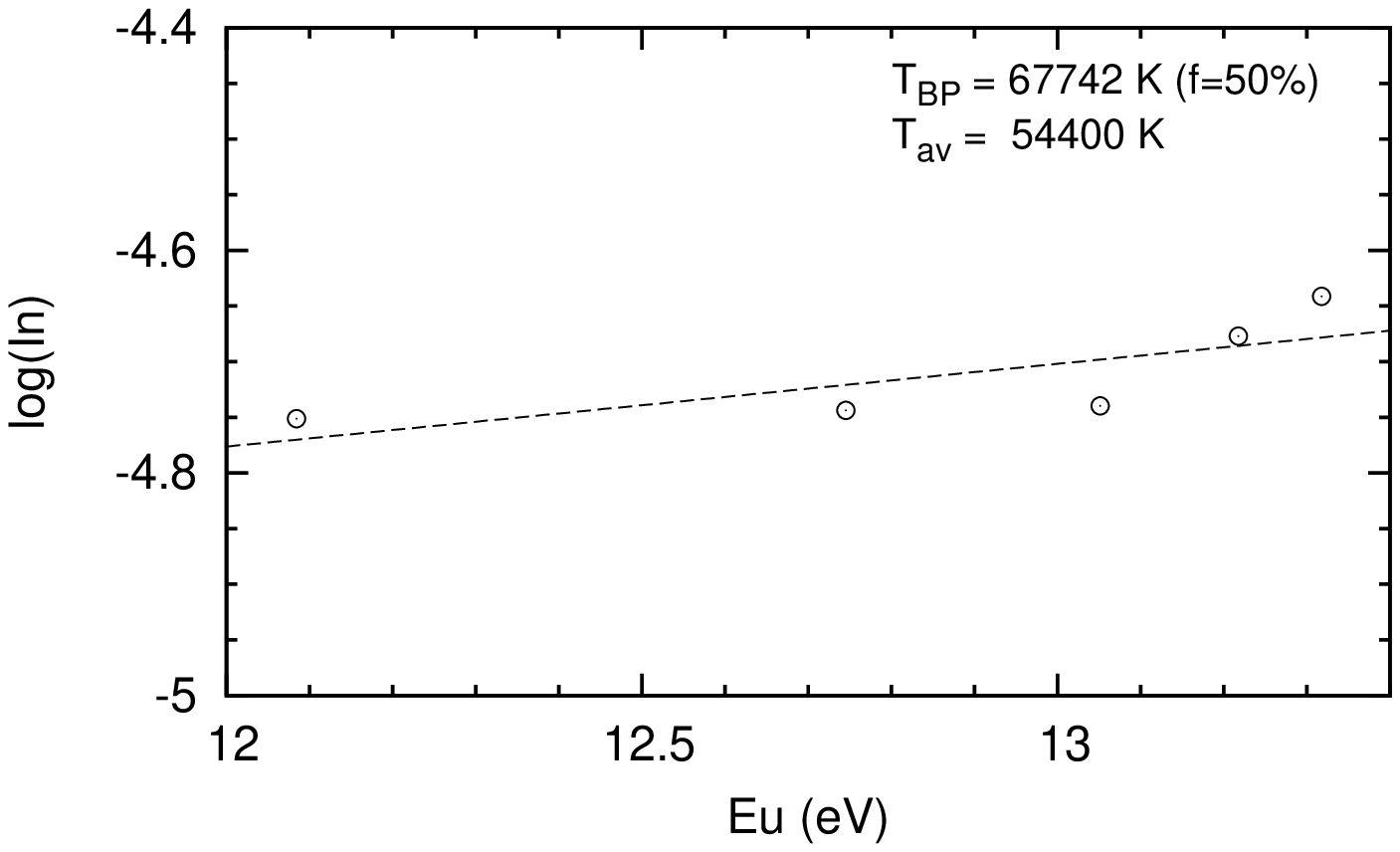}
\includegraphics[width=0.22\textwidth,angle=-90]{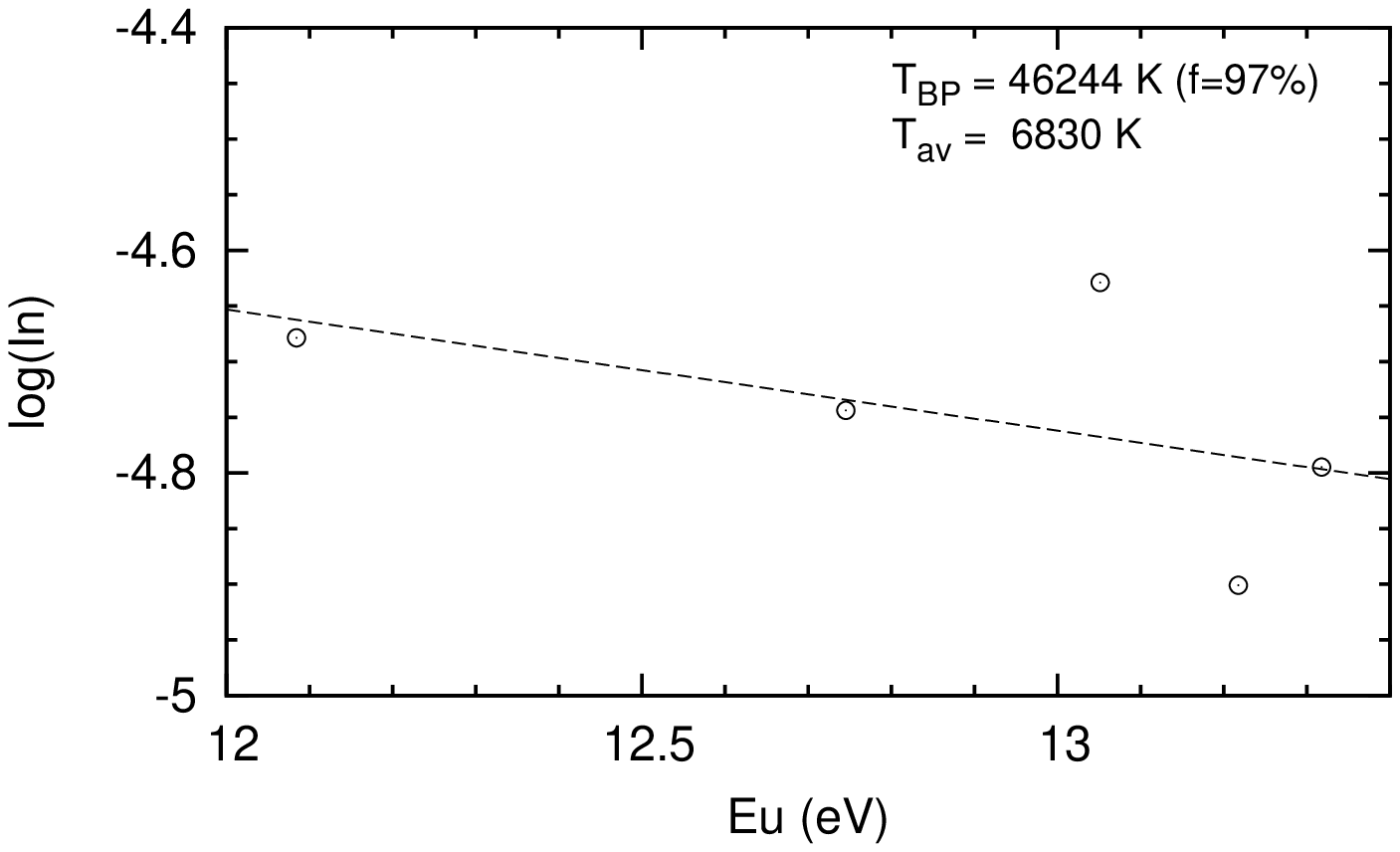}
\caption{Some examples of the BP applied on the Balmer line ratios
calculated with the CLOUDY models. The normalized intensities $F_n$
are calculated using the Balmer lines normalized on the H$\beta$
flux. In the right corner of every plot, the BP temperature $T_{\rm
BP}$, the error of the BP fit $f$ and the average temperature
$T_{\rm av}$ are given.} \label{fig00}
\end{figure}

\begin{figure*}
\centering
  \includegraphics[width=0.3\textwidth]{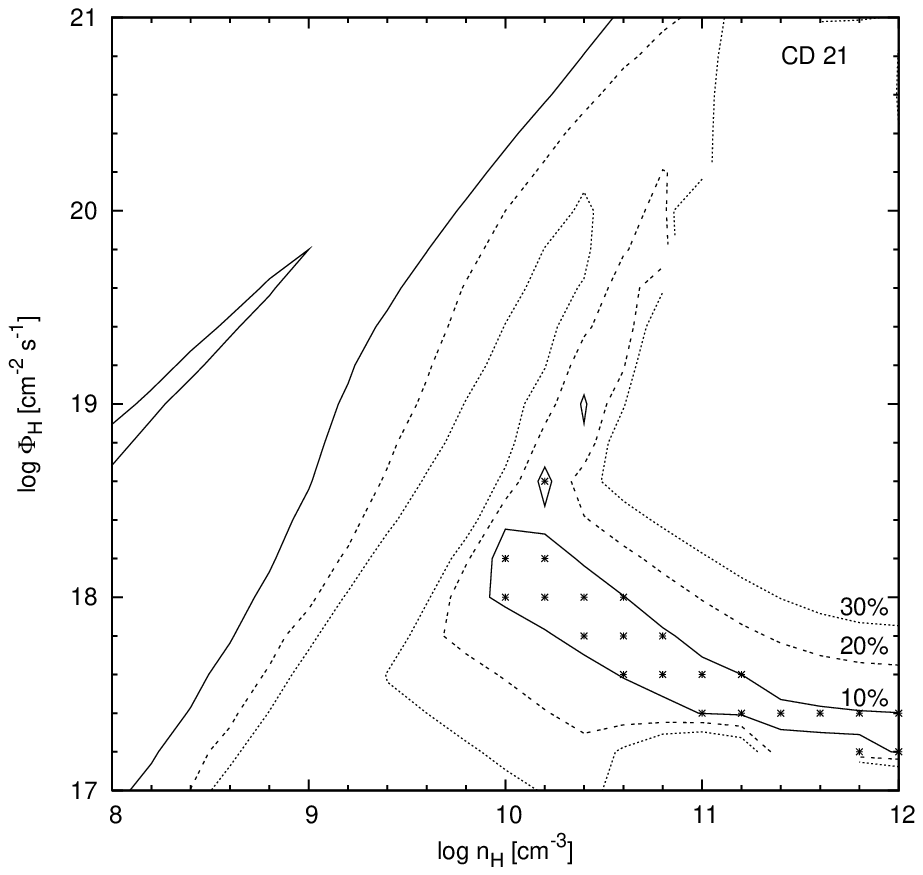}
  \includegraphics[width=0.3\textwidth]{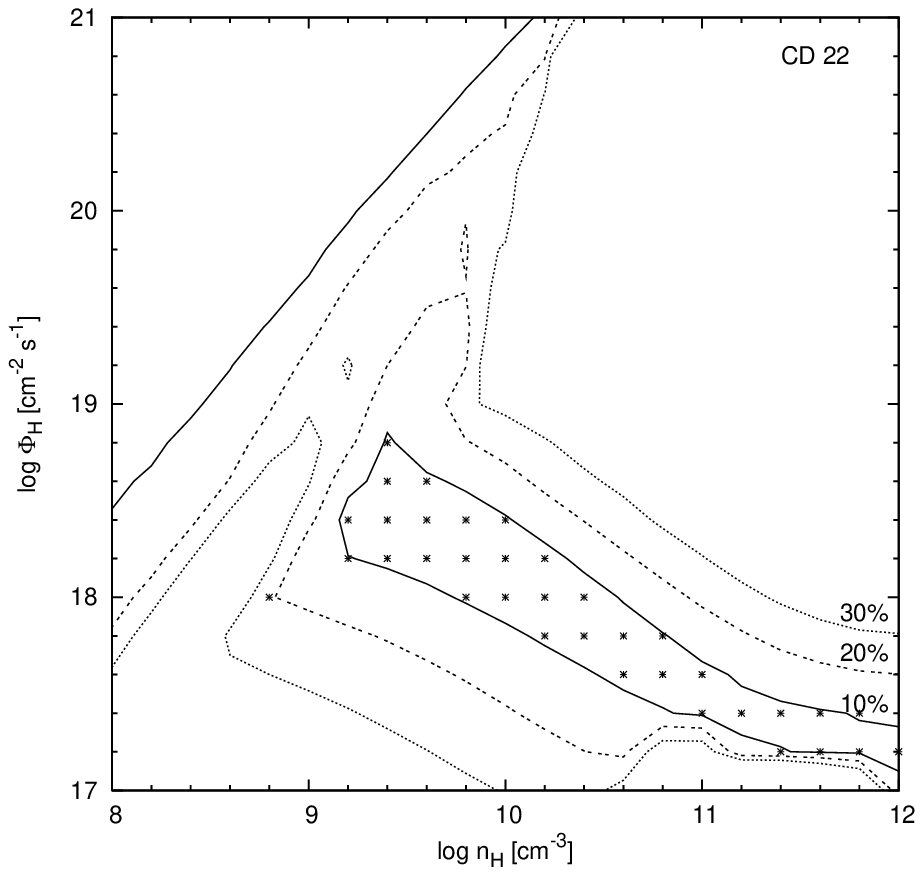}
  \includegraphics[width=0.3\textwidth]{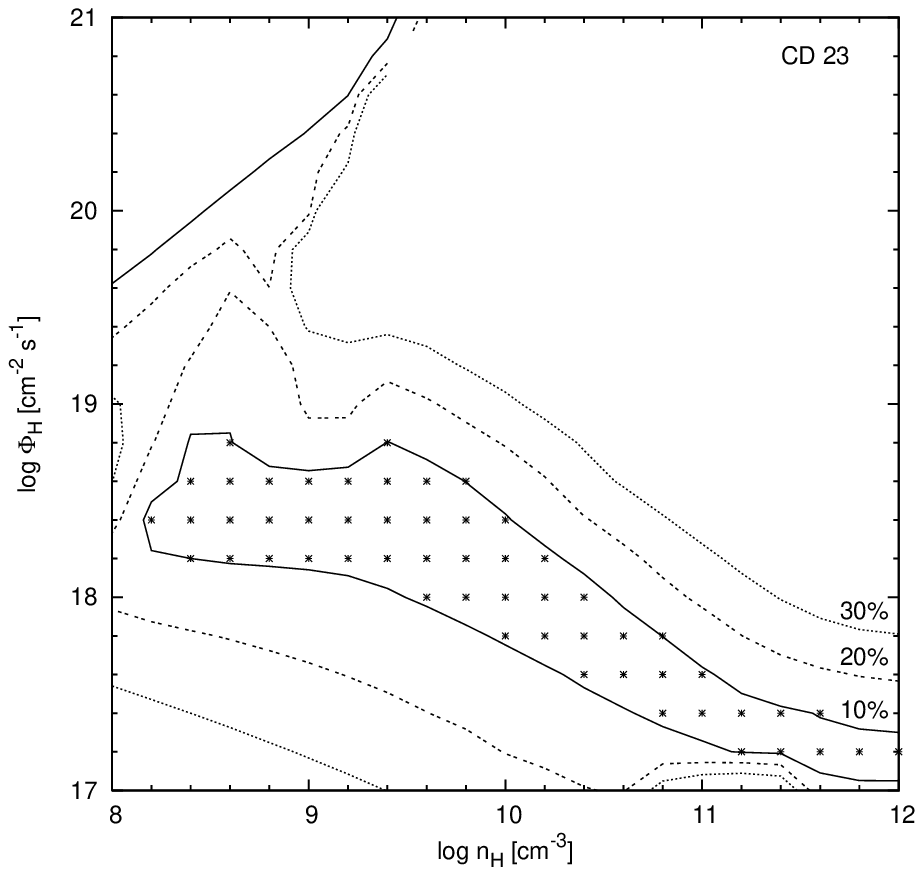}
  \includegraphics[width=0.3\textwidth]{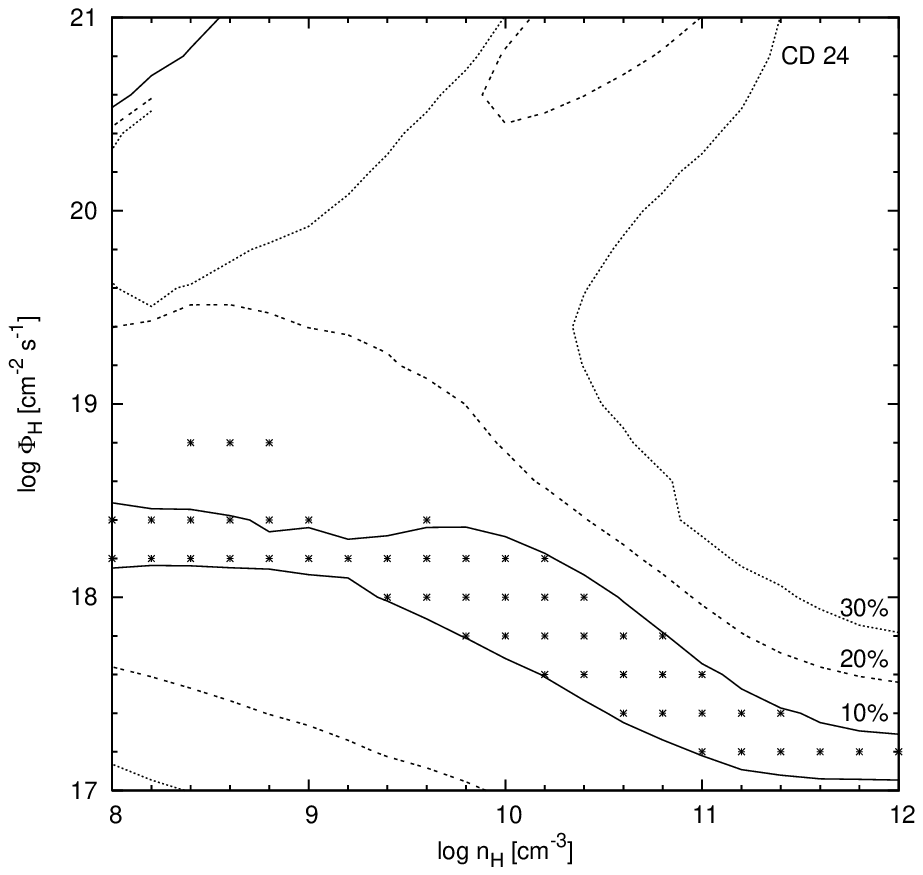}
  \includegraphics[width=0.3\textwidth]{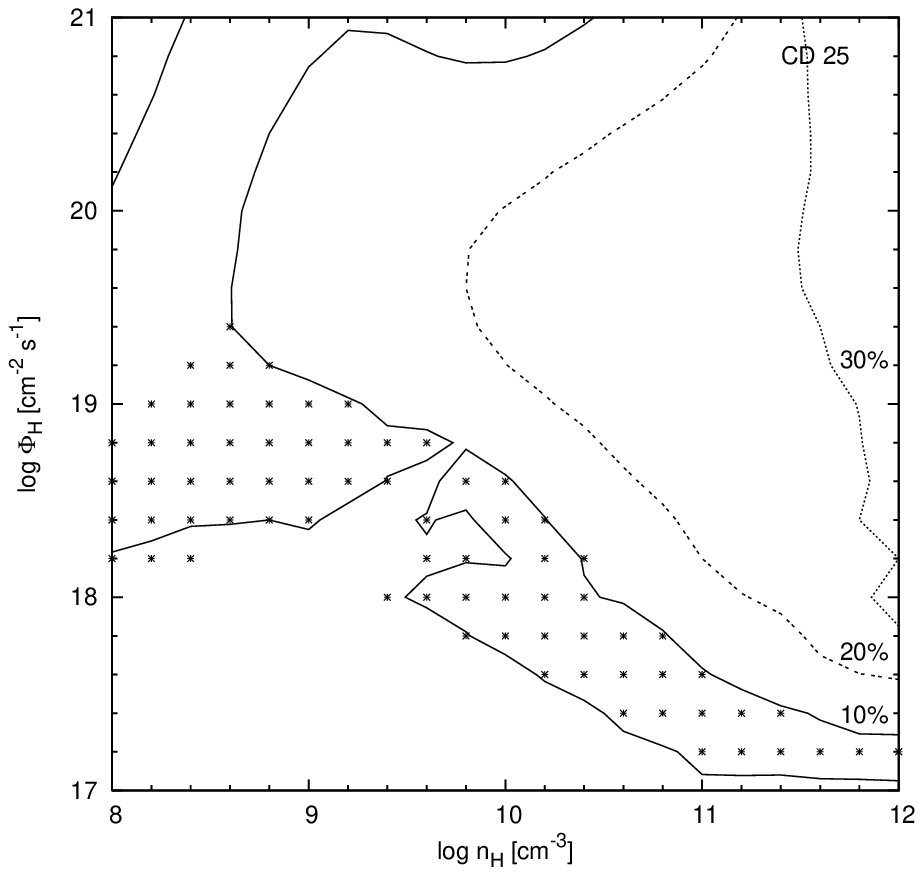}
  \caption{The error of the BP fit $f$ in the hydrogen-density vs.
ionizing flux plane for column densities $N_{\rm H} \in [10^{21} -
10^{25}] \ {\rm cm^{-2}}$ with contours inside which the BP fit
error is smaller than 10\%, 20\% and 30\%. The asterisks represent
the simulations which satisfied three defined constraints (see text
for details).}\label{fig01}
\end{figure*}

\subsection{The Boltzmann-plot method for excitation temperature diagnostics}

For the plasma of the length $\ell$ that emits along the line of
sight, assuming that the temperature and emitter density does not
vary too much, the flux $F_{ul}$ of a transition from the upper to
the lower level ($u\to l$) can be calculated as \citep{Gr97, Po03,
Po06a, Po08} $$F_{ul}={hc\over\lambda}g_{u}A_{ul}{N_0\over Z}
\exp(-E_u/kT) \ell$$ where $\lambda$ is the transition wavelength,
$g_u$ is the statistical weight of the upper level, $A_{ul}$ is
transition probability, $N_0$ is the averaged total number density
of radiating species which effectively contribute to the line flux
(which are not absorbed), $Z$ is the partition function, $E_{u}$ is
the energy of the upper level, $T$ is the averaged excitation
temperature, and $h$, $c$ and $k$ are the Planck constant, the speed
of light, and the Boltzmann constant, respectively. The additional
assumption made here is that the population of the upper level in
the transition follows the Saha-Boltzmann distribution \citep[for
more detailed derivation check][]{Po08}. From the above equation
comes the so-called Boltzmann plot (BP) that can be used to estimate
the excitation temperature $T_{\rm exc}$ in the BLR
$$\log_{10}(F_n)=\log_{10}{F_{ul}\cdot
\lambda\over{g_uA_{ul}}}=B-A{E_u}$$ where  $B$ and $A$ are BP
parameters, and $A=\log_{10}(e)/kT_{\rm exc} \approx 5040/T_{\rm
exc}$ is the temperature indicator. Therefore, for one line series
(as e.g. Balmer line series) if the population of the upper energy
states ($n\ge3$)\footnote{We note here that since the emission
deexcitation goes as $u\to l$ it is not necessary that level $l$ has
the Saha-Boltzmann distribution.}  can be described with the
Saha-Boltzmann distribution, then by applying the last equation to
that line series and obtaining the value of the parameter $A$, we
can estimate the excitation temperature of the region where these
lines are originating. We should emphasize that the additional
assumption in the BP method is that the Balmer lines are originating
in the same emitting region.

In spite of the fact that the BP method has some obvious advantages,
i.e. it is only using the measured Balmer line fluxes, that are
easily observed, to estimate the excitation temperature, one should
take into account some possible problems that could occur when using
the emission line to determine the BLR physical properties: (i) the
line profiles are usually very complex and indicate that more than
one component contribute to the total line flux, therefore one
should be aware of the multi-component BLR structure when estimating
the BEL fluxes; (ii) the Balmer lines do not have to necessarily
originate in the same region, e.g. there some indications that the
H$\alpha$ and H$\beta$ line are forming in two distinct regions
since it has been shown that the H$\beta$ line is systematically
broader than H$\alpha$ \citep{Sh08}; (iii) there are different
mechanisms contribution to the line formation. The photoionization
seems to be working well, but other heating mechanisms should be
taken into account.

\subsection{The analysis of the simulated BELs}

The first step in analyzing the BEL ratios is to apply the BP method
on the Balmer line ratios given by CLOUDY to estimate the parameter
$A$, from which we then calculate the BP temperature, i.e. the
excitation temperature ($T_{\rm BP}$ further in the text) of the
region where Balmer lines are formed. Also, from the best-fitting of
the normalized line ratios, we obtain the error of the BP fit ($f$
further in the text). A few examples of the BP applied on the Balmer
lines simulated with CLOUDY for column density of $N_{\rm H} =
10^{23} \rm cm^{-2}$ are presented in Fig.~\ref{fig00}. In many
cases a satisfactory BP fit is not obtained, i.e. $f$ has pretty
large values. This is more noticeable in the case of higher values
of the hydrogen density and ionizing-photon flux, hence we plot the
error of the BP fit $f$ in the hydrogen density vs. ionizing flux
plane for all 5 grids of models of different column density in
Fig.~\ref{fig01} (only the contours inside which $f$ is less than
10\%, 20\% and 30\% are given). From Fig.~\ref{fig01} can be seen
that if the BP method could be considered valid if $f$ is less than
10\% (eventually 20\% in the measured spectra) the parameter space
where this is valid is pretty constrained for all column densities
$N_{\rm H}$.

Out of all photoionization models, we select just those that follow
these constraints: (i) the error of the BP fit $f$ less than 10\%,
(ii) the average temperature $T_{\rm av}$ less than 20,000 K, as for
larger temperatures the Balmer line ratios are not sensitive to the
temperature changes and the BP method cannot be applied
\cite{Po06b}, and (iii) the ratio of helium lines $R$ less than 2.
Following this criteria, we constructed 5 samples of simulated BELs,
of single column density, labeled as e.g. CD21 for column density
$N_{\rm H} = 10^{21} \rm cm^{-2}$, etc. The models that satisfy
these 3 constraints are represented with asterisks in
Figure~\ref{fig01}.

\begin{figure}
\centering
\includegraphics[width=0.3\textwidth]{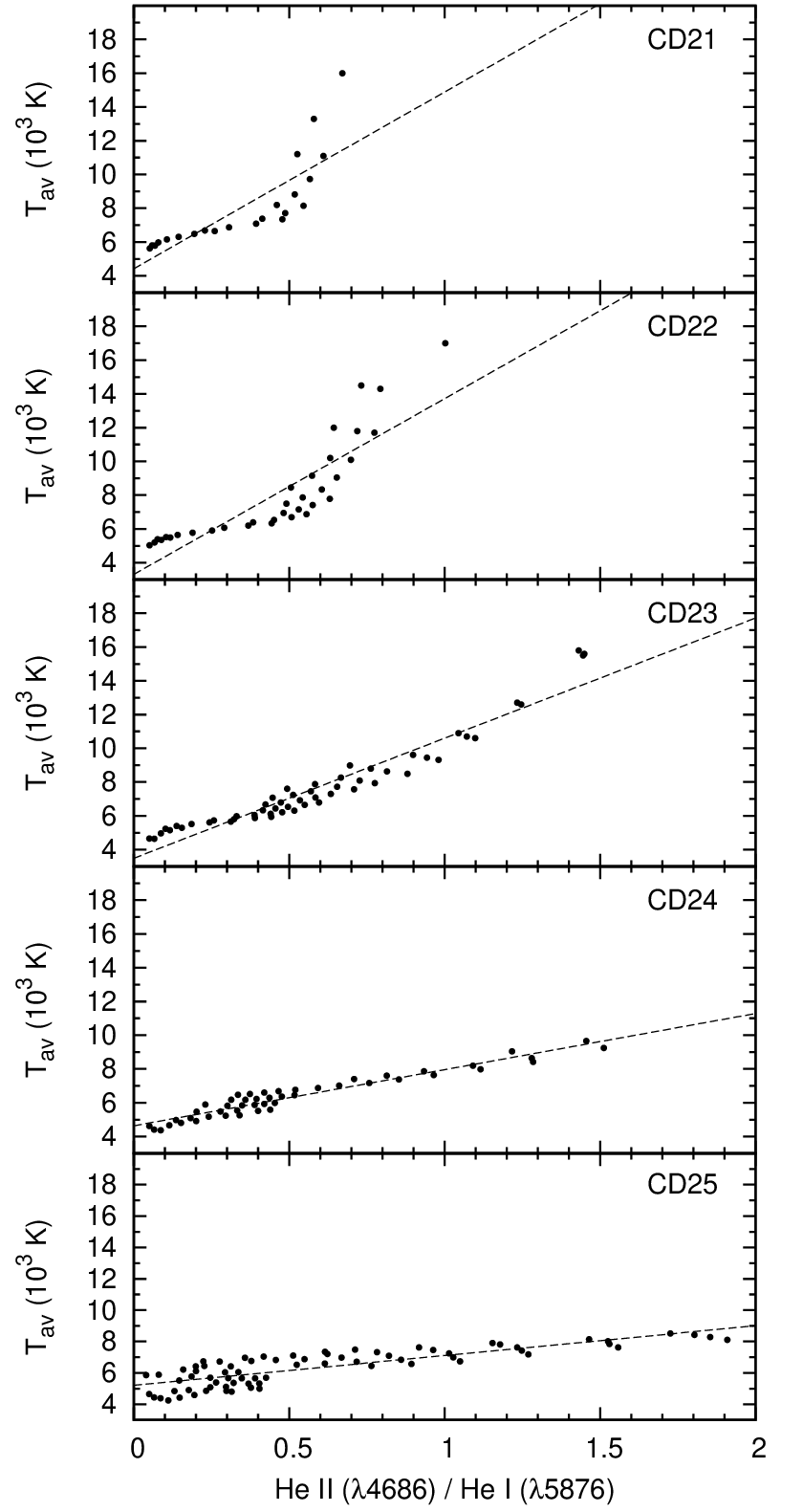}
\includegraphics[width=0.3\textwidth]{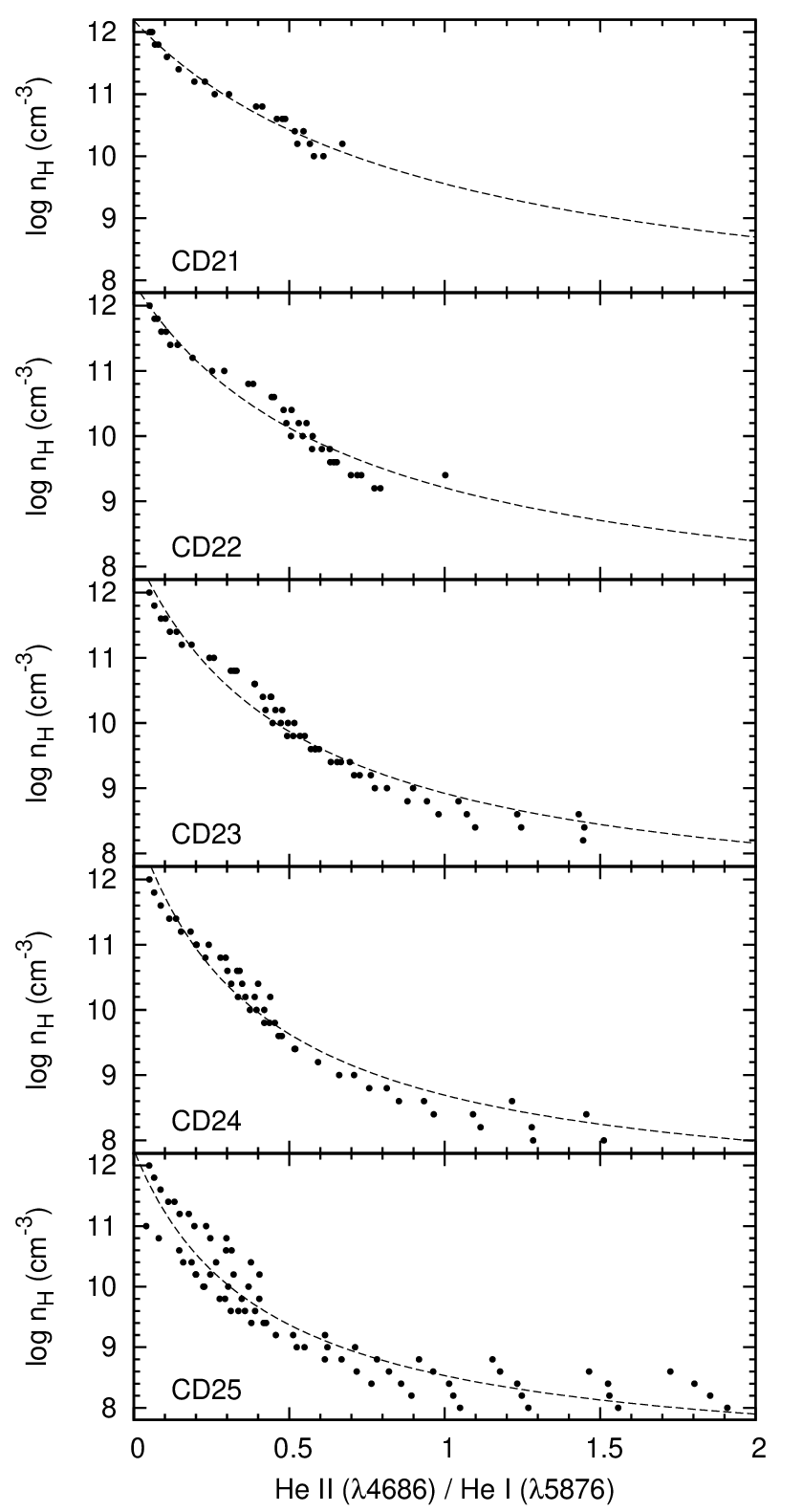}
\caption{The average temperature $T_{\rm av}$ of the emitting medium
(upper) and the hydrogen density (bottom) as the function of the
ratio $R$ of the He II $\lambda$4686 and He I $\lambda$5876 lines,
for five selected sets of simulated spectra of different column
density $N_{\rm H} = 10^{21} - 10^{25} {\rm cm^{-3}}$. The dashed
line represents the best-fitting with the function  $T_{\rm av} =
A_i + B_i \cdot R$ for the average temperature and $\log n_{\rm H} =
D_i/(C_i+R)$ for the hydrogen density. The name of the set is
specified on each plot.} \label{fig02}
\end{figure}

\subsection{The correlations between the BEL ratios and the BLR
physical properties}

We investigate in more details the five sets of simulated spectra
defined above. First, we plot the average temperature $T_{\rm av}$
of the emitting region, one of the outputs of the model, with
respect to the ratio $R$ of the helium lines for all 5 sets
(Fig.~\ref{fig02}, upper). We fit data sets with the linear function
$T_{\rm av} = A + B \cdot R$ and the best fitting results are given
in Table \ref{tab01}. Another possible connection of the BLR
physical parameters and BELs, can be obtained from the relation
between the hydrogen density and the helium line ratio $R$. We plot
the logarithm of the hydrogen density and the ratio $R$
(Fig.~\ref{fig02}, bottom), and we fitted it with the function:
$\log n_H = D/(C+R)$, where  $n_H$ is given in the units of $10^7
{\rm cm^{-3}} $. It can be seen from Fig.~\ref{fig02} (bottom) that
even when the column density changes, the same dependance of the
hydrogen density on the helium line ratio $R$ remains. The best
fitting results are given in Table \ref{tab01}.

\begin{table}
  \caption{The best-fitting parameters $A_i$ and $B_i$ of the function
$T_{\rm av} = A_i + B_i \cdot R$, and  $C_i$ and $D_i$ of the
function $\log n_{\rm H} = D_i / (C_i+ R)$ for different column
densities $N_{\rm H}$.}
  \label{tab01}
  \begin{tabular}{@{}ccccc}
  \hline
 log$N_{\rm H}$ & $A_i$ & $B_i$ &   $D_i$       & $C_i$   \\

  [cm$^{-2}$]   &   [K] &   [K] & [cm$^{-2}$] &     \\
 \hline
21 & 4422$\pm$689     &    10468$\pm$1697   &5.04$\pm$0.22   &    0.97$\pm$0.05 \\
22 & 3308$\pm$572     &    10413$\pm$1098   &3.76$\pm$0.17   &    0.70$\pm$0.04 \\
23 & 3486$\pm$200     &    7116$\pm$288     &2.91$\pm$0.10   &    0.51$\pm$0.03 \\
24 & 4634$\pm$87      &    3326$\pm$134     &2.37$\pm$0.10   &    0.40$\pm$0.03  \\
25 & 5208$\pm$119     &    1899$\pm$152     &2.16$\pm$0.12   &    0.41$\pm$0.04 \\
\hline
\end{tabular}
\end{table}

\section{The hydrogen Balmer and helium line sample}

We compare our results of the numerical simulations with the
observed data. For that we consider our measurements of well defined
sample of 90 AGN taken from the SDSS spectral database \citep{LM07},
for which the Blamer line ratios have been precisely measured and
the temperature parameter $A$ has been estimated using the BP method
applied on the Balmer line series \citep[see for details][]{LM07}.
We found that for this sample in approx 50\% of cases, the BP method
can be applied. In this analysis we consider objects that have error
of the BP fit smaller than 20\%\footnote{Even though in numerical
simulations the error was taken to be smaller than 10\%, here in
real measurements, we take larger range of fitting errors due to the
additional observational and line flux measurement errors.} and the
BP temperature smaller than 20 000 K. This reduced our sample to 48
objects.

\begin{figure}[h!]
\includegraphics[width=0.3\textwidth,angle=270]{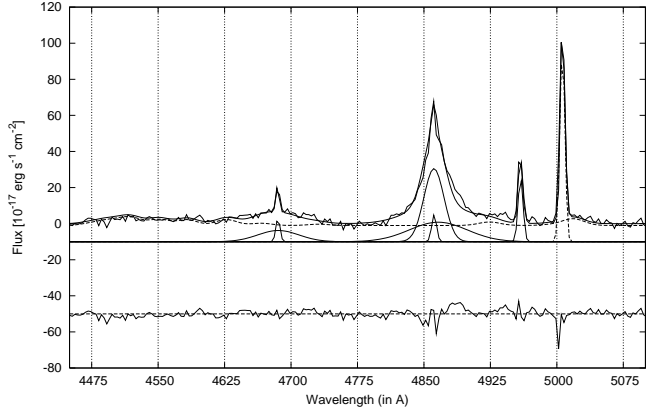}
\caption{An example of the Gaussian decomposition of the H$\beta$
region in the case of the object SDSSJ1300+5641. The Gaussian
components are shown below the observed spectrum, where the helium
line He II $\lambda$4686 is fitted with two Gaussians. The residual
spectrum is given in the bottom. The Fe II template is presented
with dashed line.} \label{fig03}
\end{figure}

For that sample, we measure the fluxes of the helium lines He II
$\lambda$4686 and He I $\lambda$5876, being particularly careful in
the case of He II $\lambda$4686 line, where there is contamination
from the Fe II multiplet and nearby H$\beta$ line. The local
continuum have been subtracted from the spectra, as well as the
contribution of the host galaxy \citep[see for details][for
details]{LM07}. For the He II $\lambda$4686 line measurements, we
perform Gaussian decomposition \citep[for details see e.g.][]{Po04}
in order to subtract the contribution of the Fe II multiplet,
H$\beta$ and narrow He II line (as an example see Fig.~\ref{fig03}).
A standard deviation is taken as an error of the line flux
measurements. The cases when helium lines could not be measured are
not considered (e.g. when lines were too noisy or the contribution
of Fe II was too strong and could not be properly subtracted), thus
our sample is reduced to 20 objects.

Using the above relations between the physical properties of the BLR
and helium line ratio $R$ (Table \ref{tab01}), we estimate the
average temperature and the hydrogen density of the BLR of these AGN
assuming different column densities. We obtain the following ranges
of physical parameters in the BLR: average temperature $T_{\rm
av}=5500 - 17600 \, \rm K$ and hydrogen density $n_{\rm H}=10^{8.3}
- 10^{11.6} \rm cm^{-3}$.

\begin{figure}
\includegraphics[width=0.28\textwidth,angle=270]{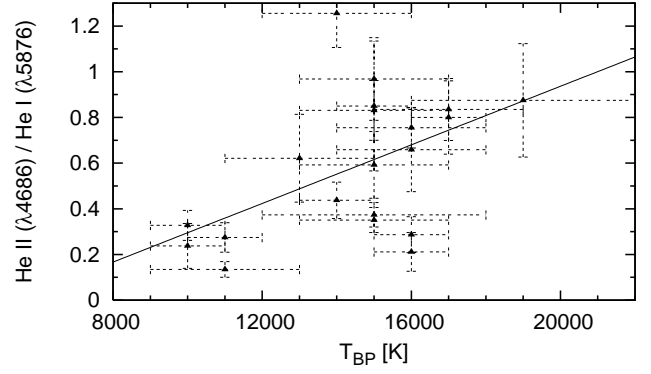}
\caption{The He II lines ratio as a function of the BP temperature
$T_{\rm BP}$  for the SDSS sample. The linear best-fitting is
presented with solid line. } \label{fig04}
\end{figure}

Moreover, in Fig.~\ref{fig04} we plot the ratio of the helium lines
as a function of the BP temperature $T_{\rm BP}$ for the SDSS
sample. As can be seen in Fig.~\ref{fig04} there is a weak
correlation (correlation coefficient $r=0.50$, $p_0=0.02321$)
between BP temperature and the He line ratio. This is also in
agreement with the BLR physics, where for higher temperatures one
should expect to have stronger He II than He I lines. We should note
here that if we exclude the point that is clearly high above the
best-fitting line (see Fig.~\ref{fig04}), the correlation
coefficient is slightly better $r=0.62$ ($p_0=0.00473$).

\section{The Fe II line emitting region}

We selected a sample of 111 AGN from Sloan Digital Sky Survey (SDSS)
according to the following criteria: high signal to noise ratio
($\mathrm{S/N}>20$), good quality of the pixels, negligible
contribution of the stellar component and a good coverage (near
uniform) of redshifts from 0 to 0.7.

\begin{figure}[h!]
\centering
\includegraphics[width=0.3\textwidth,angle=270]{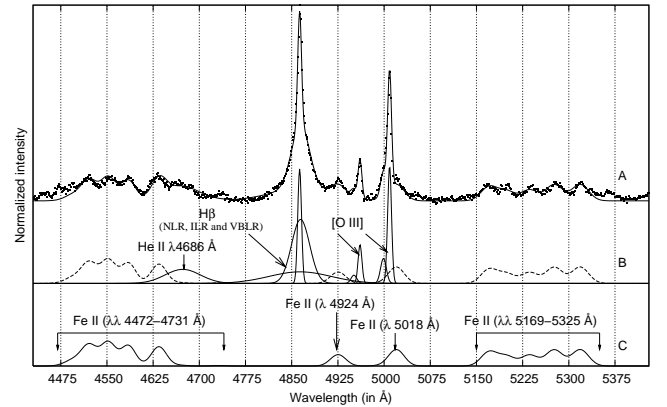}
\caption{The spectrum of SDSS J$075101.42+29174000.00$ in the
$\lambda\lambda$ 4400-5500 \AA \ range: A) observed spectra (dots)
and the best fit (solid line). B) H$\beta$ is fitted with the sum of
three Gaussians which represent emission from  NLR, ILR and VBLR.
The [O III] $\lambda\lambda$4959, 5007 \AA \ lines are fitted with
two Gaussians for each line of the doublet and He II $\lambda$4686
\AA \ is fitted with one broad Gaussian. Template of Fe II is
denoted with a dashed line, and represented also separately in panel
C (below).}\label{fig05}
\end{figure}

\begin{figure}[h!]
\centering
\includegraphics[width=0.4\textwidth]{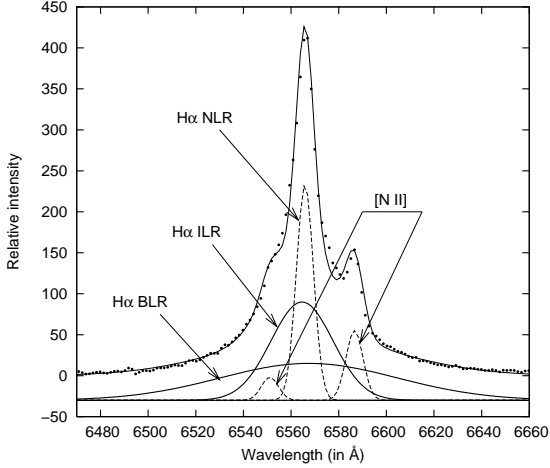}
\caption{Example of the fit of the SDSS J$075101.42+29174000.00$
spectrum in the $\lambda\lambda$ 6400-6800 \AA \ range: H$\alpha$ is
fitted with the sum of three Gaussians which represent emission from
NLR, ILR and VBLR and [N II] $\lambda\lambda$6548, 6583 \AA \ lines
are fitted with one Gaussian for each line of doublet. Narrow
emission lines (H$\alpha$ NLR and [N II]) are denoted with dashed
lines.}\label{fig06}
\end{figure}

As result, our sample contained 111 spectra, from which 58 have all
Balmer lines, and the rest 53 are without the H$\alpha$ line (due to
cosmological redshifts). After removing the influence of the
Galactic extinction and underlying continuum, the Fe II lines in the
region of the H$\beta$ line are fitted with the calculated Fe II
template. The template consist of the 33 strongest Fe II lines in
the $\lambda\lambda$ 4400-5400 \AA \ range, separated into four
groups according to the lower level of transition: $^4$F, $^6$S,
$^4$G and $^2$D1. Since all considered Fe II lines probably
originate in the same region, values of their shift and width are
assumed to be the same in one AGN, while the intensities are
different. The [O III] $\lambda\lambda$4959, 5007 \AA, He II
$\lambda$4686 \AA, [N II] $\lambda\lambda$6548, 6583 \AA \ and
Balmer lines (H$\beta$ and H$\alpha$) are fitted with a sum of
Gaussians. We assumed that Balmer lines have three components: the
narrow, intermediate and very broad, that correspond to the narrow
(NLR), intermediate (ILR) and very broad (VBLR) line region,
respectively. The NLR components in one AGN are assumed to have the
same width and shift as narrow [O III] and [N II] lines. The
examples of the fit of the H$\beta$ and H$\alpha$ line region are
shown in Fig.~\ref{fig05} and Fig.~\ref{fig06}, respectively.

\begin{figure}[h!]
\centering
\includegraphics[width=0.40\textwidth]{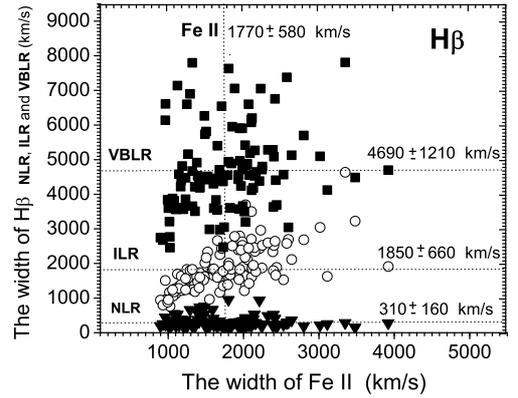}
\includegraphics[width=0.425\textwidth]{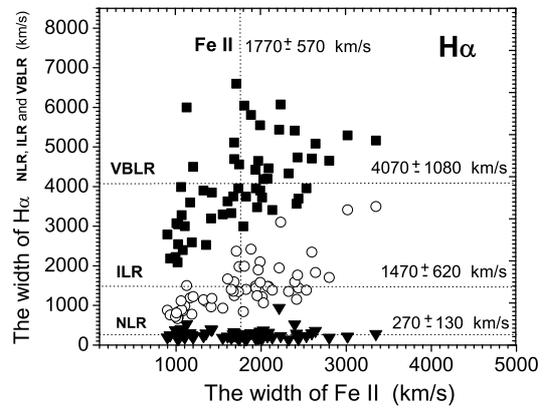}
\caption{ The widths of the Fe II lines compared to the widths of
H$\beta$ components (upper, the sample of 111 AGN) and H$\alpha$
(bottom, the sample of 58 AGN); On X-axis are widths of Fe II, and
on Y-axis are the widths of the NLR (triangles), ILR (circles) and
VBLR (squares) components of H$\beta$ (H$\alpha$). Dotted, vertical
line shows the average value of Fe II widths, while dotted,
horizontal lines show the average values of the H$\beta$ (H$\alpha$)
components. The average value of Fe II lines is the same or very
close to average width value of ILR components of H$\beta$ and
H$\alpha$.}\label{fig07}
\end{figure}

\begin{figure*}
\centering
\includegraphics[width=0.33\textwidth]{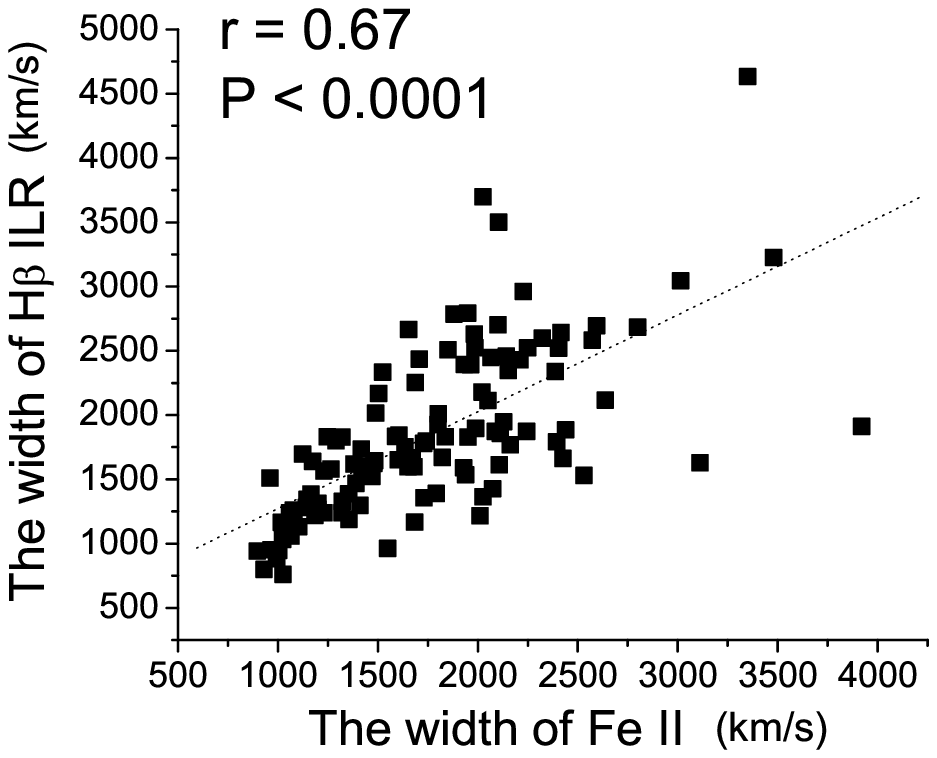}
\includegraphics[width=0.33\textwidth]{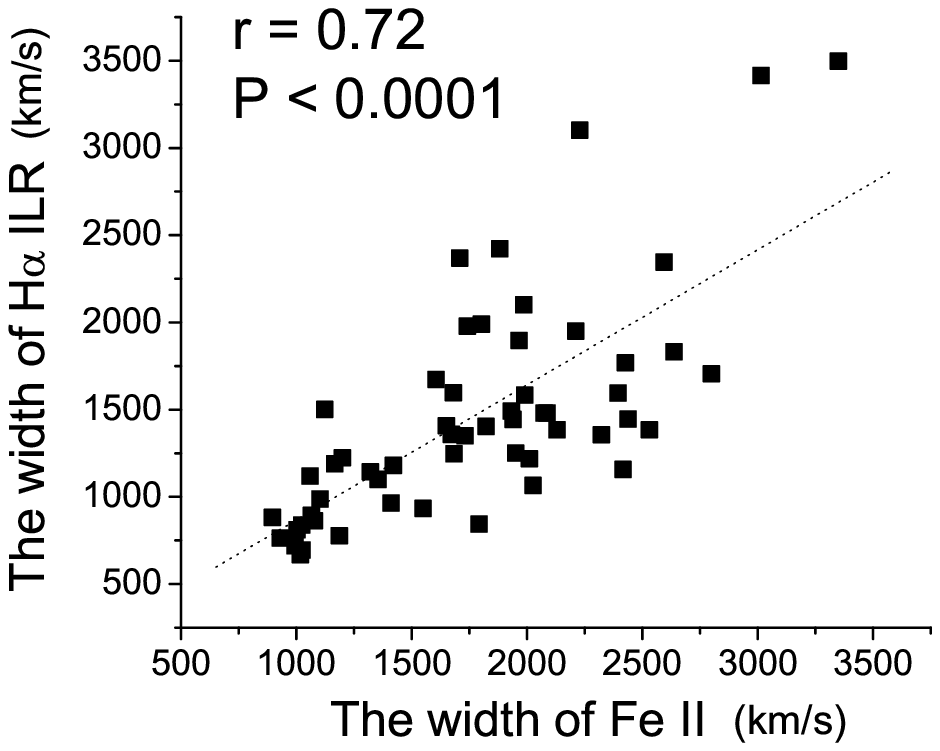}
\includegraphics[width=0.33\textwidth]{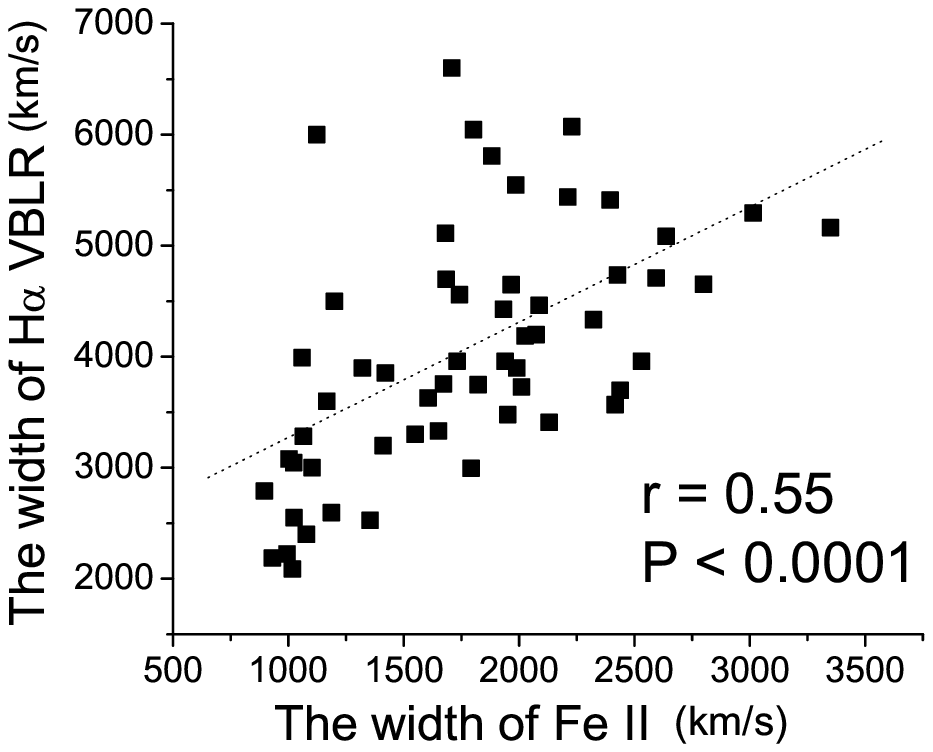}
\caption{ The correlation between the Fe II widths and widths of the
H$\beta$ ILR components (left), H$\alpha$ ILR components (middle)
and H$\alpha$ VBLR components (right). In all cases, correlations
are observed (0.67, 0.72, 0.62), indicating kinematical connection
between these emission regions.}\label{fig08}
\end{figure*}

In order to investigate the geometrical place of the Fe II emission
region in AGN, we analyzed kinematical connection between the Fe II
and Balmer emission region by comparing the line shifts and widths,
obtained from the best-fitting. We assume that the broadening of the
lines is caused by the Doppler effect from the random velocities of
emission clouds, while the shift of the line is caused by the
systemic motion of the emission gas. The results are presented in
Fig.~\ref{fig07}-\ref{fig08}.

The widths of the Fe II and H$\beta$ (upper) and H$\alpha$ (bottom)
lines are compared in Fig.~\ref{fig07}. The X-axis gives the Fe II
line widths, while the Y-axis gives the parameters of the widths of
the H$\beta$ (upper) and H$\alpha$ (bottom) components: the NLR
(triangles), ILR (circles) and VBLR (squares). Dotted lines show the
averaged values of widths: Fe II (vertical line) and H$\beta$ and
H$\alpha$ line components (horizontal lines). On the sample of the
111 AGN that contain the H$\beta$ line (upper), the average value of
Fe II widths is $1770\pm580$ km/s, while the average values for
H$\beta$ components are: $310\pm160$ km/s (NLR), $1850\pm660$ km/s
(ILR) and $4690\pm1210$ km/s (VBLR). Considering the selected sample
of 58 AGN which have the H$\alpha$ line (bottom) the averaged value
of the Fe II width is $1770\pm570$ km/s and for the H$\alpha$
components: $270\pm130$ km/s (NLR), $1470\pm620$ km/s (ILR) and
$4070\pm1080$ km/s (VBLR). It is obvious that the averaged Fe II
width ($1770$ km/s) is very close to the averaged widths of the ILR
component of both the H$\beta$ and H$\alpha$ line (1850 km/s and
1470 km/s, respectively), while the averaged widths of the NLR and
VBLR components are significantly different.

Fig.~\ref{fig08} shows the correlation between the Fe II widths and
the H$\beta$ ILR ones ($r=0.67$, $p_0<0.0001$). Considering the
H$\alpha$ ILR component, correlation is stronger, and it is
$r=0.72$, $p_0<0.0001$. We have also found that a weaker correlation
between widths of the Fe II and H$\alpha$ VBLR component ($r=0.55$,
$p_0<0.0001$) is present. Relations between the shifts of Fe II and
H$\alpha$ and H$\beta$ (NLR, ILR and VBLR components) are also
investigated. It is noticed a positive correlation with the shift of
the H$\alpha$ ILR component ($r=0.51$, $p_0<0.0001$). There is no a
significant correlation between the Fe II shift and other H$\alpha$
and H$\beta$ components.

\citet{BG92} found negative trend between the equivalent width (EW)
of Fe II and [O III] lines ($r= - 0.39$), as well as the
anticorrelation between the EW Fe II and EW ([O III])/EW (H$\beta$)
($r= - 0.53$). Thus, we tested the same relations on our sample of
111 AGN. We found slightly lower correlations then the previous
authors: for EW Fe II vs. EW [O III] the correlation coefficient is
$r= -0.26$ ($p_0=0.007$) and for EW Fe II vs. EW ([O III])/EW
(H$\beta$) it is $r= -0.37$ ($p_0<0.0001$).

Then, we divided our initial sample of 111 AGN (0$<$z$<$0.7) in 6
subsamples, which are made by gradually removing the objects with
low redshift in 6 steps. Subsamples contain object with redshifts in
intervals: 0.1$<$z$<$0.7, 0.2$<$z$<$0.7, 0.3$<$z$<$0.7,
0.4$<$z$<$0.7, 0.5$<$z$<$0.7 and 0.6$<$z$<$0.7. Relation between EW
Fe II and EW [O III] (also EW [O III]/EW H$\beta$) are analyzed
separately for each subsample (see Table~\ref{tab02}). We found that
negative trend (r = - 0.37, P$<$0.0001) observed between EW Fe II
and EW [O III]/EW H$\beta$ in initial sample (0$<$z$<$0.7), progress
to significant anticorrelation (r$\sim$ - 0.70, P$<$0.0001) as we
are removing low-redshift objects. The redshift dependance of the EW
Fe II vs. EW [O III] anticorrelation may indicate some cosmological
repercussion of the Fe II emission in AGN or dependence on
luminosity.

\begin{table}
  \caption{Correlations between equivalent widths (EW) of the Fe II
and [O III] ([O III]/H$\beta$) lines for subsamples within different
redshift intervals. Relation is fitted with function $Y=A+B\cdot X$.
The coefficient of correlations r, the corresponding measure of the
significance of correlations $p_0$, as well as the A and B
coefficients, are shown in the table. In the last column, the number
of objects in each subsample is given.  }
  \label{tab02}
  \begin{tabular}{@{}ccccc}
  \hline
 vs. &   & EW [O III] &   EW [O III]/EW H$\beta$  & No   \\

 \hline
 EW FeII &$r$&-0.26& -0.37&109\\
(0$<$z$<$0.7)&$p_0$ &0.007 &$<0.0001$& \\
\hline
EW FeII  &$r$&-0.30& -0.40&92\\
(0.1$<$z$<$0.7)&$p_0$ &0.004  &$<0.0001$& \\
\hline
EW FeII &$r$&-0.42& -0.50&74\\
(0.2$<$z$<$0.7)&$p_0$ & 2.2E-4 & $<0.0001$&\\
\hline
EW FeII &$r$&-0.44& -0.56&58\\
(0.3$<$z$<$0.7)&$p_0$ &  6.0E-4& $<0.0001$&\\
\hline
EW FeII &$r$&-0.56& -0.62&43\\
(0.4$<$z$<$0.7)&$p_0$ & $<0.0001$ &$<0.0001$&\\
\hline
EW FeII &$r$&-0.64& -0.70&28\\
(0.5$<$z$<$0.7)&$p_0$ &2.37E-4 & $<0.0001$&\\
\hline
EW FeII &$r$&-0.68& -0.72&14\\
(0.6$<$z$<$0.7)&$p_0$ &0.007 & 0.004&\\
\hline
\end{tabular}
\end{table}

\section{Results and discussion}

In this progress report we give some results of our recent
investigations of different broad line parameters (the flux ratios,
EW, FWHM, etc.) in order to probe the physics of the ELR, i.e. the
hydrogen Balmer lines (H$\alpha$ to H$\varepsilon$), the helium
lines from two subsequent ionization levels (He II $\lambda$4686 and
He I $\lambda$5876) and the strongest Fe II lines in the wavelength
interval $\lambda\lambda 4400-5400 \AA$.

From the analysis of the hydrogen Balmer lines, we can say that the
BP method gives valid results if the error of the BP fit $f$ is less
than 10\% (eventually 20\% in the measured spectra). The parameter
space of CLOUDY simulations where $f<10\%$ is well constrained and
in a similar range for different column densities $N_{\rm H}$
(Fig.~\ref{fig01}). This area of parameters contains lower ionizing
fluxes and higher hydrogen densities, but depending on the column
density the area increases keeping the same trend between the
$n_{\rm H}$ and $\Phi_{\rm H}$, Therefore, for those parameters the
BP method could be applied for the estimation of the excitation
temperature. This indicates that for some physical conditions, even
if we have the photoionization as the main heating process in the
BLR, the hydrogen Balmer lines are produce in such way that they
obey the Saha-Boltzmann equation, i.e. the BLR plasma is at least in
the Partial Local Thermodynamical Equilibrium.

The physical parameters of the simulations for which the BP method
could be applied ($f<10\%$) follow some relations even when the
column density changes. There is a linear relation between the
average temperature $T_{\rm av}$ of the emitting region and helium
lines ratio $R$: $T_{\rm av} = A + B \cdot R$ (Fig.~\ref{fig03},
Table~\ref{tab01}). Also, the hydrogen density depends on the helium
line ratio $R$ as: $\log n_H = D/(C+R)$, where  $n_H$ is given in
the units of $10^7 {\rm cm^{-3}} $ (Figure~\ref{fig03},
Table~\ref{tab01}). For lower column densities ($N_{\rm H}=10^{21} -
10^{22} \rm cm^{-2}$) this linear trend for $T_{\rm av}$ seems to be
broken after some value of $R$ (Fig.~\ref{fig03}, upper), therefore
the results obtained for this column densities should be taken with
caution. The ranges of the average temperature and hydrogen density,
obtained when these relations are applied to the observed sample of
AGN taken from the SDSS database, are in good agreement with the
previous estimates of the physical conditions in the BLR
\citep{OF06}. Having in mind problems given in \S 2.1 (i.e.
different emitting regions of helium and Balmer lines or the
multicomponent origin of BELs) the relations given in
Table~\ref{tab01} could be use as a rough estimate of the BLR
physical parameters from direct measurements.

On  the other hand, from the analysis of the Fe II emission, we
found the significant correlation between the Fe II and H$\alpha$,
H$\beta$ ILR widths which indicate the ILR origin of the Fe II
emission. Also, the Fe II emission region is mainly characterized
with random velocity of $\sim$1800 km/s, that corresponds to the ILR
origin. This result is in favor to the results obtained by
\citet{Po04}, and with the recent findings that the optical Fe II
line forming region seems to not be the same as the BLR \citep{Ku08,
Hu08, Po09}. Moreover, we found that the degree of anticorrelation
of EW Fe II - EW [O III] (EW [O III]/H$\beta$) is redshift
dependent, i.e. anticorrelation increases by selection of the sample
with high redshift objects.

\section{Conclusions}

We used here the emission lines to estimate the physical parameters
of the BLR and properties of the Fe II line forming region. From our
investigations we can outline the following conclusions:
\begin{enumerate}[(i)]

\item In the case of the pure photoionized plasma, there can exist the
conditions that the excitation of the hydrogen Balmer lines is
sensitive to the temperature. In this case, the He II/He I line
ratio can be used for the estimates of the averaged temperature and
hydrogen density. This can be very useful for determining the
physical properties of the BLR, and for a small sample of AGN we
found that the BLR temperatures are ranging from $\sim$ 5000 K to
18000 K, and hydrogen densities from 10$^{8}$ cm$^{-3}$ to 10$^{12}$
cm$^{-3}$. These values of the temperature and density are expected
in the BLR plasma \citep{OF06}.

\item The optical Fe II emission is likely coming from a region that
corresponds to the ILR. Also, there is a weak correlation with the
VBLR component of Balmer lines indicating that a fraction emitted in
the Fe II lines can originate also in the BLR. On the other hand,
the correlation between the EW FeII and EW [OIII] (EW[OIII]/EW
H$\beta$) discovered by \citet{BG92} seems to be sensitive on
redshift (or luminosity) of AGN.

\end{enumerate}

Finally, the BLR physics (and geometry) still remains an open
question, and we hope that with this work we approached some
conclusions which will help us in understanding the BLR physics.

\vskip 5mm

\hskip 2.8cm Acknowledgments

\vskip 3mm

D.I. would like to thank to the Department of Astronomy of the
University of Padova in Italy for their hospitality. This work was
supported by the Ministry of Science of Serbia through the project
Astrophysical Spectroscopy of Extragalactic Objects (\#146002).

Funding for the SDSS and SDSS-II has been provided by the Alfred P.
Sloan Foundation, the Participating Institutions, the National
Science Foundation, the U.S. Department of Energy, the National
Aeronautics and Space Administration, the Japanese Monbukagakusho,
the Max Planck Society, and the Higher Education Funding Council for
England. The SDSS is managed by the Astrophysical Research
Consortium (ARC) for the Participating Institutions. The SDSS Web
Site is http://www.sdss.org/. This research has made use of NASA's
Astrophysics Data System.

\bibliographystyle{elsarticle-harv}




\end{document}